\documentclass{aa}
\usepackage{graphicx}
\usepackage{txfonts}
\usepackage{natbib,twoopt}
\usepackage[breaklinks=true, colorlinks=true,urlcolor=blue,citecolor=blue,linkcolor=blue]{hyperref} 
\bibpunct{(}{)}{;}{a}{}{,}             
\makeatletter
  \newcommandtwoopt{\citeads}[3][][]{\href{http://adsabs.harvard.edu/abs/#3}%
    {\def\hyper@linkstart##1##2{}%
     \let\hyper@linkend\@empty\citealp[#1][#2]{#3}}}
  \newcommandtwoopt{\citepads}[3][][]{\href{http://adsabs.harvard.edu/abs/#3}%
    {\def\hyper@linkstart##1##2{}%
     \let\hyper@linkend\@empty\citep[#1][#2]{#3}}}
  \newcommandtwoopt{\citetads}[3][][]{\href{http://adsabs.harvard.edu/abs/#3}%
    {\def\hyper@linkstart##1##2{}%
     \let\hyper@linkend\@empty\citet[#1][#2]{#3}}}
  \newcommandtwoopt{\citeyearads}[3][][]%
    {\href{http://adsabs.harvard.edu/abs/#3}
    {\def\hyper@linkstart##1##2{}%
     \let\hyper@linkend\@empty\citeyear[#1][#2]{#3}}}
\makeatother

\begin{document}

\title{Shape-based clustering of synthetic Stokes profiles using $k$-means and $k$-Shape}

\author{Thore E. Moe
	\inst{1,2}
	\and Tiago M.D. Pereira
	\inst{1,2}
	\and Flavio Calvo
	\inst{3}
	\and Jorrit Leenaarts
	\inst{3}
	}
\institute{Rosseland Centre for Solar Physics, University of Oslo, P.O. Box 1029 Blindern, NO--0315 Oslo, Norway
\and
Institute of Theoretical Astrophysics, University of Oslo, P.O. Box 1029 Blindern, NO--0315 Oslo, Norway
\and
Institute for Solar Physics, Dept. of Astronomy, Stockholm University, AlbaNova University Centre, 10691 Stockholm, Sweden} 
\date{}

\abstract 
{The shapes of Stokes profiles contain much information about the atmospheric conditions that produced them. However, a variety of different atmospheric structures can produce very similar profiles. Thus, it is important for proper interpretation of observations to have a good understanding of how the shapes of Stokes profiles depend on the underlying atmosphere. An excellent tool in this regard is forward modeling, i.e. computing and studying synthetic spectra from realistic simulations of the solar atmosphere. Modern simulations routinely produce several hundred thousand spectral profiles per snapshot. With such numbers, it becomes necessary to use automated procedures in order to organize the profiles according to their shape. Here we illustrate the use of two complementary methods, $k$-means and $k$-Shape, to cluster similarly shaped profiles, and demonstrate how the resulting clusters can be combined with knowledge of the simulation's atmosphere to interpret spectral shapes.} 
{We aim to showcase the use of clustering analysis for forward modeling. In particular we wish to introduce the $k$-Shape clustering method to the solar physics community as a complement to the well-known $k$-means method.} 
{We generate synthetic Stokes profiles for the \ion{Ca}{II} 854.2 nm line using the Multi3D code from a Bifrost simulation snapshot. We then apply the $k$-means and $k$-Shape clustering techniques to group the profiles together according to their shape, and investigate the within-group correlations of temperature, line-of-sight velocity and line-of-sight magnetic field strengths.} 
{We show and compare the classes of profile shapes we retrieve from applying both $k$-means and $k$-Shape to our synthetic intensity spectra. We then show the structure of the underlying atmosphere for two particular classes of profile shapes retrieved by the clustering, and demonstrate how this leads to an interpretation for the formation of those profile shapes. Furthermore, we apply both methods to the subset of our profiles containing the strongest Stokes $V$ signals, and demonstrate how $k$-Shape can be qualitatively better than $k$-means at retrieving complex profile shapes when using a small number of clusters.} 
{} 


\maketitle

\section{Introduction}
\label{sec:intro}

Forward modeling of the solar atmosphere is a very useful tool for understanding the relative importance of atmospheric components in the formation of polarized spectra, thereby guiding interpretations of observations. By computing synthetic Stokes profiles from realistic 3D radiative magnetohydrodynamic (rMHD) simulations, one can directly compare a particular spectral signature with the full state of the atmosphere that produced it \citep[see e.g.][and others in the same series]{2013ApJ...772...90L, 2013ApJ...772...89L, 2013ApJ...778..143P}. Modern simulations routinely contain several hundred thousand pixels, with each pixel giving rise to a set of Stokes profiles. Depending on the spatial resolution of the numerical model, and the spectral resolution considered for the synthesis, these profiles can be quite complex; often exhibiting more complicated behavior than what is typically resolved in real observations. It is obviously not feasible to analyze the formation of so many profiles one by one, nor is it practical to manually sort them into groups according to their features. Rather, some automated procedure must be used to organize the profiles in a meaningful manner for further human analysis.

One way of reducing the number of individual profiles into more manageable collections is the use of clustering techniques like $k$-means \citep{1956_Steinhaus, 1967_Macqueen}. $k$-means has seen extensive use in solar and stellar physics, for examples see \citet{2000ApJ...532.1215S, 2007ApJ...663.1386P, 2011A&A...530A..14V, 2018ApJ...861...62P, 2019ApJ...875L..18S, 2019A&A...631L...5B, 2020A&A...640A..71K, 2020A&A...641L...5J, 2021ApJ...922..137W,  2021A&A...655A..28N, 2021A&A...651A.112B, 2021A&A...647A.147B, 2022A&A...657A.132K, 2022A&A...668A.153M, 2022arXiv221109103S}.
Apart from $k$-means, other clustering methods have also been used on solar spectra, for instance the $t$-distributed Stochastic Neighbor Embedding employed by \citep{2021ApJ...907...54V}.
The purposes of the clustering vary from identifying and studying the observational signatures of particular physical processes and features, to reducing the spatial dimensionality of data-sets for inversions, to statistical characterizations of observations. Relatively little explored, however, is the application of clustering techniques in a forward modeling context, one notable exception being \citet{2005A&A...442.1059K}. In this paper we aim to address that issue, applying the $k$-means method to \ion{Ca}{II} 854.2 nm  Stokes $I$ and Stokes $V$ profiles generated from a Bifrost \citep{2011A&A...531A.154G} snapshot using the Multi3D radiative transfer code \citep{2009ASPC..415...87L}, which has been extended (Calvo \& Leenaarts (in prep.)) to include polarization, accounting for the Zeeman effect. We focus on the shapes of the Stokes profiles, aiming to illustrate what different classes of shapes do, or do not, tell us about the underlying atmospheric conditions.

\begin{figure}
        \centering
                \includegraphics[width=9cm]{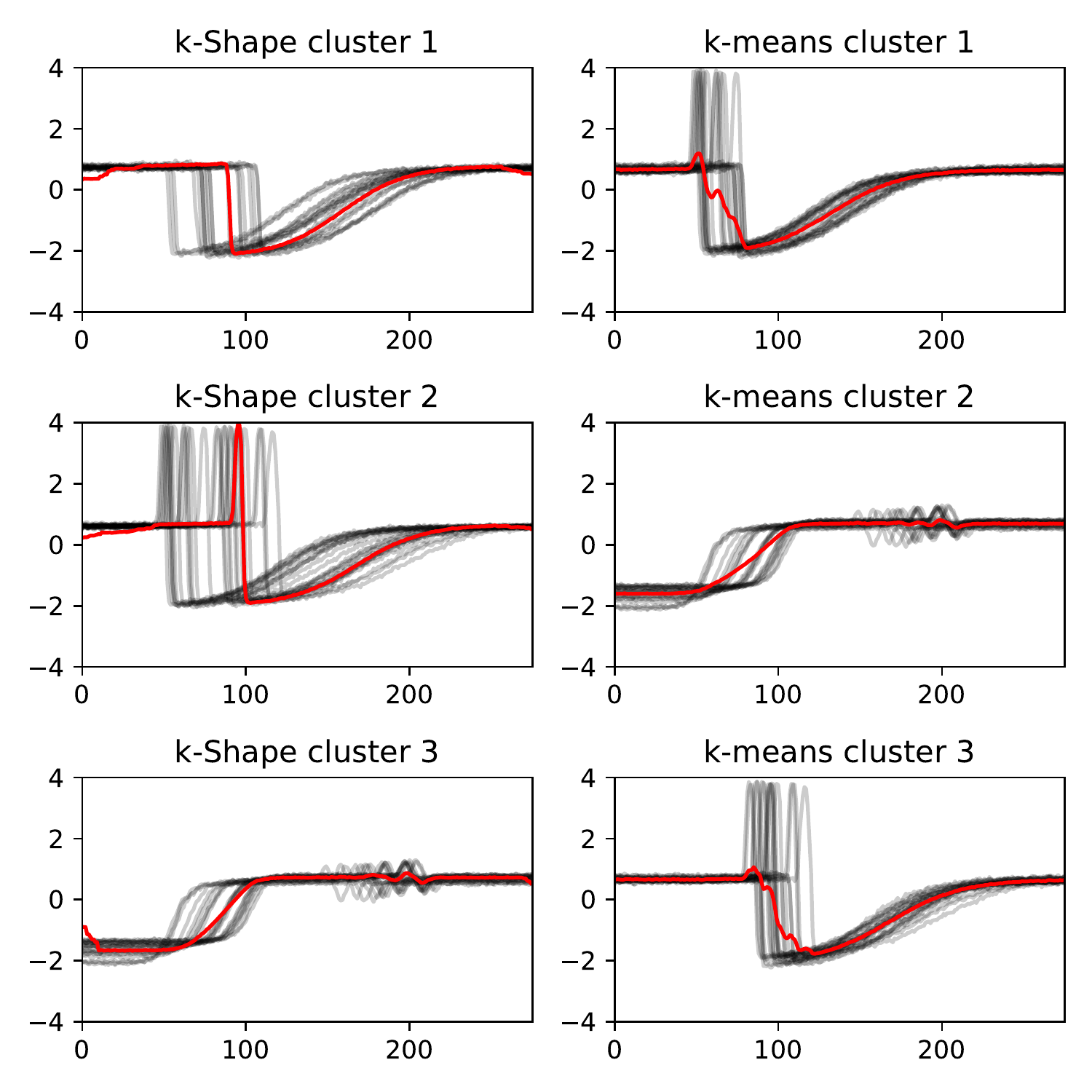}
                        \caption{Example showing how $k$-Shape (left) and $k$-means (right) partition a dataset with three distinct signal shapes. While $k$-Shape recovers the three distinct classes of shapes, $k$-means mixes the class containing a peak and a drop with the class containing only a drop. This illustration is adapted from the documentation of the tslearn-library. \protect\footnotemark}
                        \label{fig:ks_vs_km}
\end{figure}

While $k$-means is a fast and robust clustering technique, it does not directly cluster profiles based on their shapes. It works by minimizing the sum of within-cluster Euclidean distances between profiles, which can potentially lead to distinctly different shapes appearing in the same cluster as demonstrated in Fig. \ref{fig:ks_vs_km}. Or, for instance, two Doppler-shifted spectral profiles with the otherwise same exact shape can be put into separate clusters. Furthermore, the centroid, or `representative profile' (RP), of a cluster is given as the mean of the profiles belonging to the cluster, which in some cases can give a poor representation of the typical profile shapes in the cluster. Of course, increasing the number of clusters can mitigate this problem, but at the cost of the interpretability, which is the main point of the kind of forward modeling we seek to undertake in this paper.

A relatively fast clustering method that is inherently shape-based is the $k$-Shape method of \citet{10.1145/2723372.2737793}. Though originally developed for use on time-series, the method is quite general and we apply it here to the case of Stokes profiles with the obvious substitution of the time axis for a wavelength axis. A feauture of $k$-Shape is that the clustering is largely independent of Doppler-shifts, which can be beneficial or detrimental depending on the intended usage case. By ignoring Doppler-shifts and using a different measure of similarity than $k$-means, the profiles are matched more directly according to their similarity in actual shape, rather than being matched according to a combination of shape and wavelength position. Furthermore, as the centroid computation is rather different from the one in $k$-means, the RP's are much more prototypical of the clustered profiles. The cost, of course, is that all absolute velocity-information is not considered in the clustering.

\footnotetext{\url{https://tslearn.readthedocs.io/en/stable/auto_examples/clustering/plot_kshape.html\#sphx-glr-auto-examples-clustering-plot-kshape-py}}

\section{Methods}
\label{sec:methods}

\subsection{Generating synthetic profiles}

We generated our synthetic spectra from the 23 km resolution atmospheric model described in \citet{2022A&A...662A..80M}. This is a Bifrost model \citep{2011A&A...531A.154G} with a magnetic field configuration constructed to resemble a coronal hole. The model has $512\times512\times512$ grid points, spanning roughly 12 Mm in the horizontal directions and going from $z=-2.5$~Mm below up to $z=8$~Mm above the solar surface. The horizontal spacing of the grid points is uniform, resulting in a horizontal resolution of 23~$\mathrm{km}\,\mathrm{pix}^{-1}$. 

We used an extension (Calvo \& Leenaarts (in prep.)) of the Multi3D code \citep{2009ASPC..415...87L} with polarimetric capabilities to produce 3D full Stokes profiles of the \ion{Ca}{II}~854.2 nm line accounting for the Zeeman effect. As 3D computations are immensely expensive we cut the bottom 112 grid points, corresponding to below $-0.4$~Mm beneath the surface, under the assumption that these are too deep to affect the formation of our line of interest. Furthermore, we neglected to include the effects of partial frequency redistribution (PRD) and isotopic splitting. The obtained synthetic profiles were normalized by the nearby continuum, meaning each profile was divided by the Stokes I value of the reddest wavelength in the synthesis at approximately $\lambda_0 + 0.95$ nm, and interpolated to 100 equidistant wavelength points in the range $\lambda_0 \pm 0.05$ nm, where $\lambda_0$ denotes the central wavelength of the line. We performed this interpolation in order to give equal weight to all parts of the profile when clustering since the original wavelength grid used in the synthesis is non-equidistant.

\subsection{$k$-means clustering}

The most common clustering technique for spectral profiles is $k$-means clustering. The full set of profiles is divided into $k$ clusters of similarly shaped profiles, where the number $k$ must be chosen at the outset. The measure of similarity is the Euclidean distance between profiles; that is, the distance between two profiles is the sum over wavelengths of the squared difference in their amplitudes:
\begin{equation}
\label{eq:euclidean_distance}
distance = \sum_i (I_1(\lambda_i) - I_2(\lambda_i))^2,
\end{equation}
where $I(\lambda_i)$ denotes the amplitude of the profile at each wavelength point $\lambda_i$.
Each cluster has a centroid, and the goal is to assign the profiles to the $k$ clusters in such a way that the sum of distances between all profiles and their nearest centroid (often called the inertia) is minimized. Algorithmically, $k$-means performs the following steps: 
\begin{enumerate}
\item Initialize $k$ centroids, one for each cluster.
\item Assign each profile to the cluster with the closest centroid.
\item Recompute the centroids as the mean (for each wavelength) of the profiles belonging to the cluster.
\item Repeat 2. and 3. until no profile changes cluster, a fixed number of iterations has been performed, or until the total inertia no longer changes above a set tolerance.
\end{enumerate}

It should be noted that the convergence of the $k$-means algorithm does not guarantee that a global minimum has been found. Therefore it is common to re-initialize the clustering a predefined number of times, keeping the result with lowest inertia. 

In this paper, we have used the $k$-means implementation of scikit-learn \citep{scikit-learn}, employing the $k$-Means++ initialization \citep{10.5555/1283383.1283494} for selecting better initial cluster centroids.

\subsection{$k$-Shape clustering}

As the name implies, $k$-Shape \citep{10.1145/2723372.2737793}, is designed to perform a clustering into $k$ clusters of distinct shape. While the general idea is similar to $k$-means, it uses a different metric for the distance between profiles; as well as another method for computing the cluster centroids. The distance metric is based on shifting the profiles across each other and computing the cross-correlation for each possible shift. Consider two profiles $I_1$ and $I_2$, defined on $m$ wavelength points, written in the form of vectors:
\begin{equation}
\label{eq:intensities_as_vector}
\vec{I_1} = I_1(\lambda_1),I_1(\lambda_2),...,I_1(\lambda_m),\;\; \vec{I_2} = I_2(\lambda_1),I_2(\lambda_2),...,I_2(\lambda_m).
\end{equation}
The cross-correlation sequence between these two profiles, $CC_w(\vec{I_1},\vec{I_2})$, is defined as:
\begin{equation}
\label{eq:def_CC}
CC_w(\vec{I_1},\vec{I_2}) = R_{w-m}(\vec{I_1},\vec{I_2}), \quad w \in \{1,2,\dots,2m-1\},
\end{equation}
where
\begin{equation}
\label{eq:def_R}
R_k(\vec{I_1},\vec{I_2}) = \begin{cases} 
\sum_{l=1}^{m-k}{I_1(\lambda_{l+k}) \cdot I_2(\lambda_l)}, \quad k \geq 0
\\
R_{-k}(\vec{I_1},\vec{I_2}), \quad k < 0.
\end{cases}
\end{equation}

Thus, the sequence $CC_w(\vec{I_1},\vec{I_2})$ contains the cross-correlation value for each of the $2m-1$ possible shifts of the profiles relative to each other; essentially a sequence of the vector dot products between zero-padded $\vec{I_1}$ and $\vec{I_2}$ for each possible overlapping shift of the profiles. Normalizing the cross-correlation sequence (corresponding to dividing by the Euclidean norm of both profiles):
\begin{equation}
\label{eq:normalized_cc}
NCC_c = \frac{CC_w(\vec{I_1},\vec{I_2})}{\sqrt{R_0(\vec{I_1},\vec{I_1}) \cdot R_0(\vec{I_2},\vec{I_2})}},
\end{equation}
results in a number between $-1$ and $1$ for each entry in the sequence, where $-1$ signifies perfect anti-correlation and $1$ signifies perfect correlation between the profiles. Selecting the entry with the largest cross-correlation value then gives the shape-based distance between two profiles as:
\begin{equation}
\label{eq:kshape_distance}
distance = 1 - \max_w \bigg(\frac{CC_w(\vec{I_1},\vec{I_2})}{\sqrt{R_0(\vec{I_1},\vec{I_1}) \cdot R_0(\vec{I_2},\vec{I_2})}}\bigg),
\end{equation}
which is bounded between 0 and 2. 

As in $k$-means, each profile is assigned to the closest centroid in terms of distance, and the cluster centroid is recomputed. In $k$-Shape, however, the refinement of the cluster centroids is done by reformulating the minimization of within-cluster distances as a maximization of a Rayleigh quotient calculation; for details see the original paper \citep{10.1145/2723372.2737793}. It should, however, be remarked that the $k$-Shape method assumes that the profiles have been $z$-normalized, meaning each profile has zero mean, and unity standard deviation:
\begin{equation}
\label{eq:z_normalization}
\vec{I_1}^\prime = \frac{\vec{I_1} - \mu_1}{\sigma_1},
\end{equation}
where $\mu_1$ and $\sigma_1$ is, respectively, the mean and the standard deviation of the profile over the $m$ wavelengths considered. This assumption is not strictly necessary, as the method can be modified to work with other data-normalizations. However, the original authors found the z-normalization to work best in their tests and it is beyond the scope of our current work to re-implement and evaluate the method for other normalizations.

We used the $k$-Shape implementation from the tslearn library \citep{JMLR:v21:20-091}, with some simple modifications to make it run in parallel. Even so, the $k$-Shape method is significantly slower than the $k$-means implementation of scikit-learn. In one example case, using $k=100$ clusters for $512 \times 512$ profiles with 100 wavelength points, one run of $k$-Shapes without re-initializations took roughly $2.7$ hours, while a $k$-means run with 10 re-initializations took about 5 minutes, both on the same 32-core workstation. It should be noted that in the tslearn implementation of $k$-Shape, $k$ single profiles are randomly chosen as the initial cluster centroids. In the original paper \citep{10.1145/2723372.2737793}, the initialization is done by randomly distributing all profiles among $k$ clusters

\section{Results}
\label{sec:results}

\subsection{Overview}

\begin{figure*}
        \centering \hspace*{-5mm}
                \includegraphics[width=19.5cm]{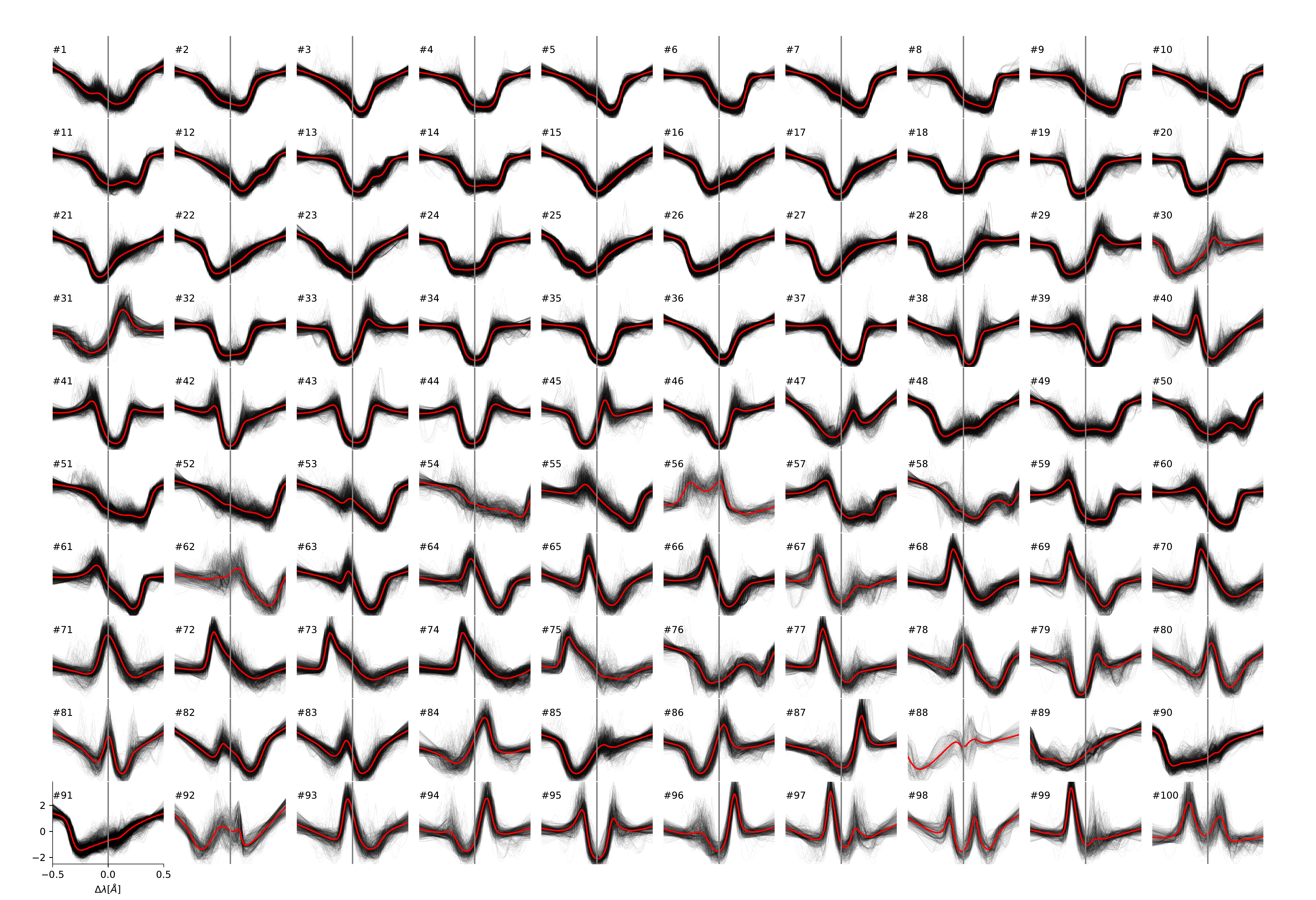}
                        \caption{$k$-means clusters for synthetic \ion{Ca}{II} 854.2~nm intensity profiles, using 100 clusters on $z$-normalized profiles. The red line is the cluster centroid profile (average), while the black lines are all individual profiles assigned to each cluster. The grey line is a visual aid that denotes the position of $\lambda_0$}
                        \label{fig:kmie_clusters}
\end{figure*}

\begin{figure*}
        \centering \hspace*{-5mm}
                \includegraphics[width=19.5cm]{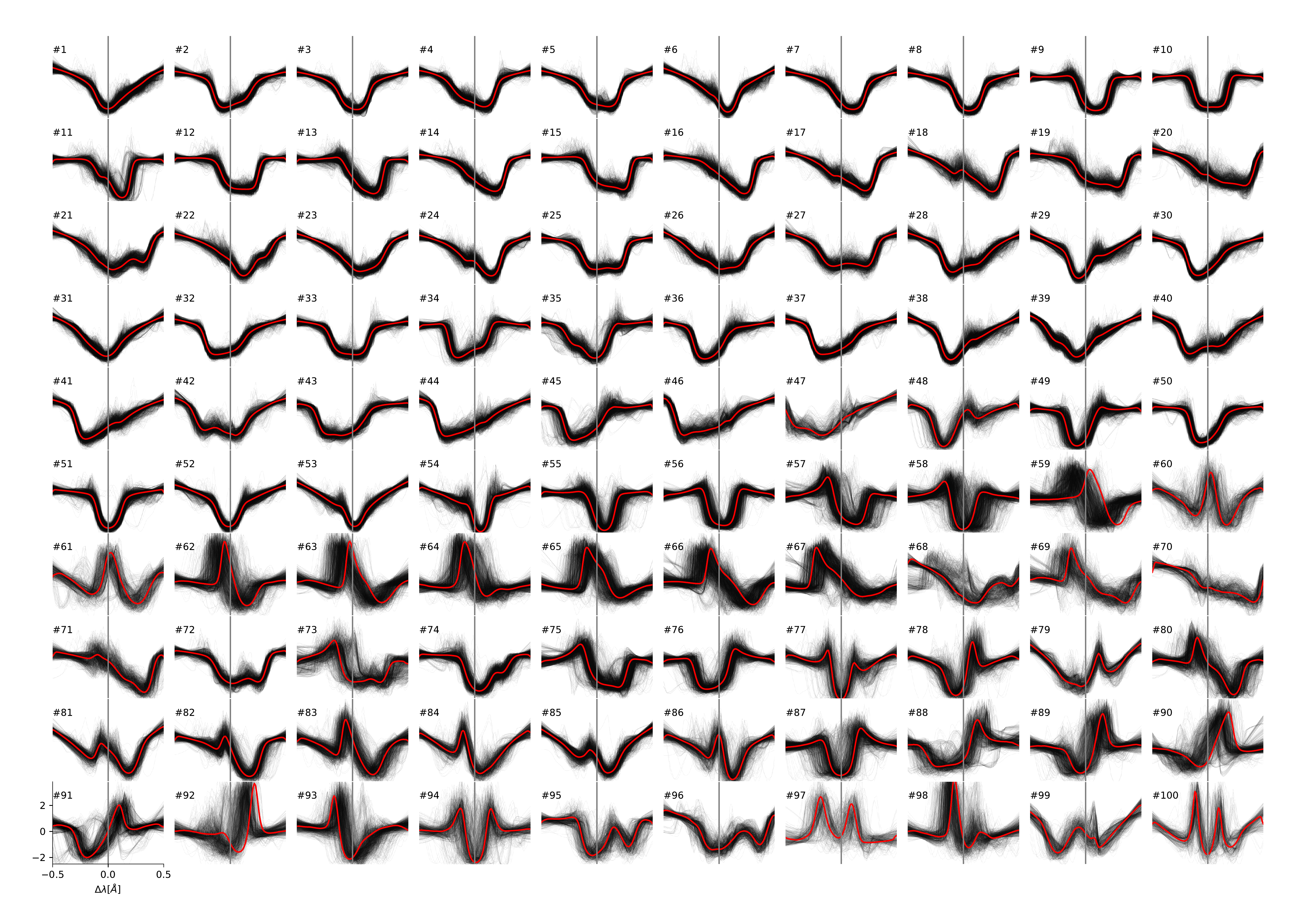}

                        \caption{$k$-Shape clusters for synthetic \ion{Ca}{II} 854.2~nm intensity profiles, using 100 clusters on $z$-normalized profiles. The red line is the cluster centroid from $k$-Shape, and the black lines are all individual profiles assigned to the cluster. The grey line is a visual aid that denotes the position of $\lambda_0$}
                        \label{fig:ksie_clusters}
\end{figure*}

Our intention was to illustrate and compare the use of both $k$-Shape and $k$-means for clustering synthetic profiles according to their shape, and subsequently how the resulting clusters can reveal correlations between the typical profile shapes in a cluster and the particular structure of the underlying atmosphere these profiles emerge from. We therefore begin by presenting and discussing the clustering of the intensity profiles in Sec. \ref{sec:intensity_clusters}, before we perform a detailed examination of two particular profile shapes retrieved by the clustering in Sec. \ref{sec:cbg_like} and Sec. \ref{sec:double_peaked}.

As the $k$-Shape method assumes that its input profiles are $z$-normalized, we used the same normalization for the $k$-means method in order to do a fair comparison. This turned out to be a reasonable approach for the synthetic intensity profiles, as they have signal values in the same general range. However, the polarized components of the Stokes vector can vary vastly in amplitude, so the $z$-normalization can cause tiny signals to appear misleadingly large compared to stronger signals as the amplitude is given in units of the per-profile standard deviation. We have therefore focused mostly on the intensity profiles, though we did perform a clustering of the very strongest Stokes $V$ signals (those with a signal exceeding $0.5\%$ of the nearby continuum intensity), which we will discuss in Sec. \ref{sec:strong_V}.

\subsection{Clustering the intensity profiles}
\label{sec:intensity_clusters}

We clustered the synthetic intensity profiles into $k=100$ clusters using both $k$-means and $k$-Shape, the resulting clusters are shown in Fig. \ref{fig:kmie_clusters} and Fig. \ref{fig:ksie_clusters} respectively. The choice of 100 clusters was made after some experimentation, as a reasonable trade-off between the two opposing considerations of accuracy and human interpretability. The $k$-means method was run with 10 re-initializations, while the $k$-Shape method was run with a single initialization due to being around two orders of magnitude slower. We have tested $k$-Shape with 10 re-initializations, which yielded qualitatively very similar results to the single initialization run. We therefore elected to use the single initialization run in order to compare the methods for somewhat more similar runtimes.

The first observation we can make is that both clustering techniques seem to recover a similar variety of different profile shapes. These range from typical absorption profiles (e.g. \#35 in Fig. \ref{fig:kmie_clusters}, \#52 in Fig. \ref{fig:ksie_clusters}), through increasingly strongly skewed absorption profiles (e.g. \#9 and \#37 in Fig. \ref{fig:kmie_clusters}, \#30 and \#44 in Fig. \ref{fig:ksie_clusters}), to more complicated profiles, including double-peaked profiles (e.g. \#98 and \#100 in Fig. \ref{fig:kmie_clusters}, \#97 and \#100 in Fig. \ref{fig:ksie_clusters}), asymmetric emission profiles (e.g. \#73 in Fig. \ref{fig:kmie_clusters}, \#64 in Fig. \ref{fig:ksie_clusters}) and multi-lobed profiles (e.g. \#81 in Fig. \ref{fig:kmie_clusters} or \#84 in Fig. \ref{fig:ksie_clusters}). 

The clustering appears to be reasonably tight, and in both methods there are several clusters showing very similar shapes, i.e. there is more than one cluster per `family' of shapes. Encouragingly, both clustering methods seem to recover all the same types of cluster `families', e.g. several clusters with similar asymmetric emission peaks or double peaks show up in both clusterings, though there is obviously not a one-to-one correspondence between individual clusters across the methods. Conversely, at first glance there do not seem to be clusters with very distinct shapes found only with one method compared to the other. The most unique-looking clusters are perhaps \#56 and \#88 in Fig. \ref{fig:kmie_clusters}, but even these find quite similar counterparts in \#97 and \#47 in Fig. \ref{fig:ksie_clusters}. This gives us some confidence that our choice of 100 clusters reasonably covers the range of typical profile shapes.

A second observation we can make, is how the retrieved clusters do differ between the methods. The $k$-Shape groupings demonstrate the method's insensitivity to Doppler-shifts, especially the clusters containing the asymmetric emission peaks (e.g. \#63, \#64, \#65, in Fig. \ref{fig:ksie_clusters}) show the same shape at different shifts grouped together. Conversely, $k$-means splits these into different clusters (e.g. \#72, \#73, \#74 in Fig. \ref{fig:kmie_clusters}) according to their Doppler shifts. The fact that both methods retrieve the same `families', but differently distributed over the clusters, can be beneficial for analysis, as we will see in Sect. \ref{sec:cbg_like}. With such a stereoscopic view of the underlying atmospheres it becomes easier to discern by inspection which atmospheric parameters are important and which are incidental for the formation of the particular profile shapes. In particular, $k$-Shape's insensitivity to Doppler shifts contrasted with $k$-means sensitivity to them, allows one to better discern which atmospheric behaviors are correlated solely with the shape of the profile, as opposed to being correlated to the combination of shape and Doppler-shift.

A third observation relates to how and where the methods perform poorly, in terms of profiles not being a good fit for their assigned clusters. As mentioned, cluster \#56 in $k$-means does not seem to be well captured by $k$-Shape. It turns out that most of the profiles from this cluster are assigned to \#68 and \#73 in Fig. \ref{fig:ksie_clusters}. These profiles are on the whole quite different from their assigned $k$-Shape centroids, but when the profiles and the centroids are shifted drastically across each other, the overlapping parts agree sufficiently for them to be grouped together. As $k$-Shape computes all possible shifts, it may occasionally find large shifts (and thereby a large clipping of the signal) to be the least bad option, leading to such apparently poor assignments. That type of signal clipping does not happen with $k$-means.

On the other hand, the $k$-means clusters appear to have issues distinguishing profiles where there is a large difference in signal strength over a narrow wavelength region. For instance, the $k$-means cluster \#79 in Fig. \ref{fig:kmie_clusters} turns out to be a mix of profiles with enhanced shoulders on either the right side or on both sides of the line core, as well as some with only a weakly enhanced right shoulder followed by a second absorption feature to the right. In the $k$-Shape clustering vast majority of these profiles are assigned to \#77, \#78, \#94 and \#95 in Fig. \ref{fig:ksie_clusters}.

To summarize, neither method performs ideally, in the sense that both have clusters where some members that are rather poorly represented by the centroids. The obvious way to improve the fidelity of the clusters is to increase the number of clusters, or possibly do more re-initializations. However, the methods seem to complement each other, each to an extent balancing out the others weaknesses, and are useful as starting points for human analysis.

\subsection{CBG-like profiles}
\label{sec:cbg_like}

\begin{figure*}
        \centering
                \includegraphics[width=18cm]{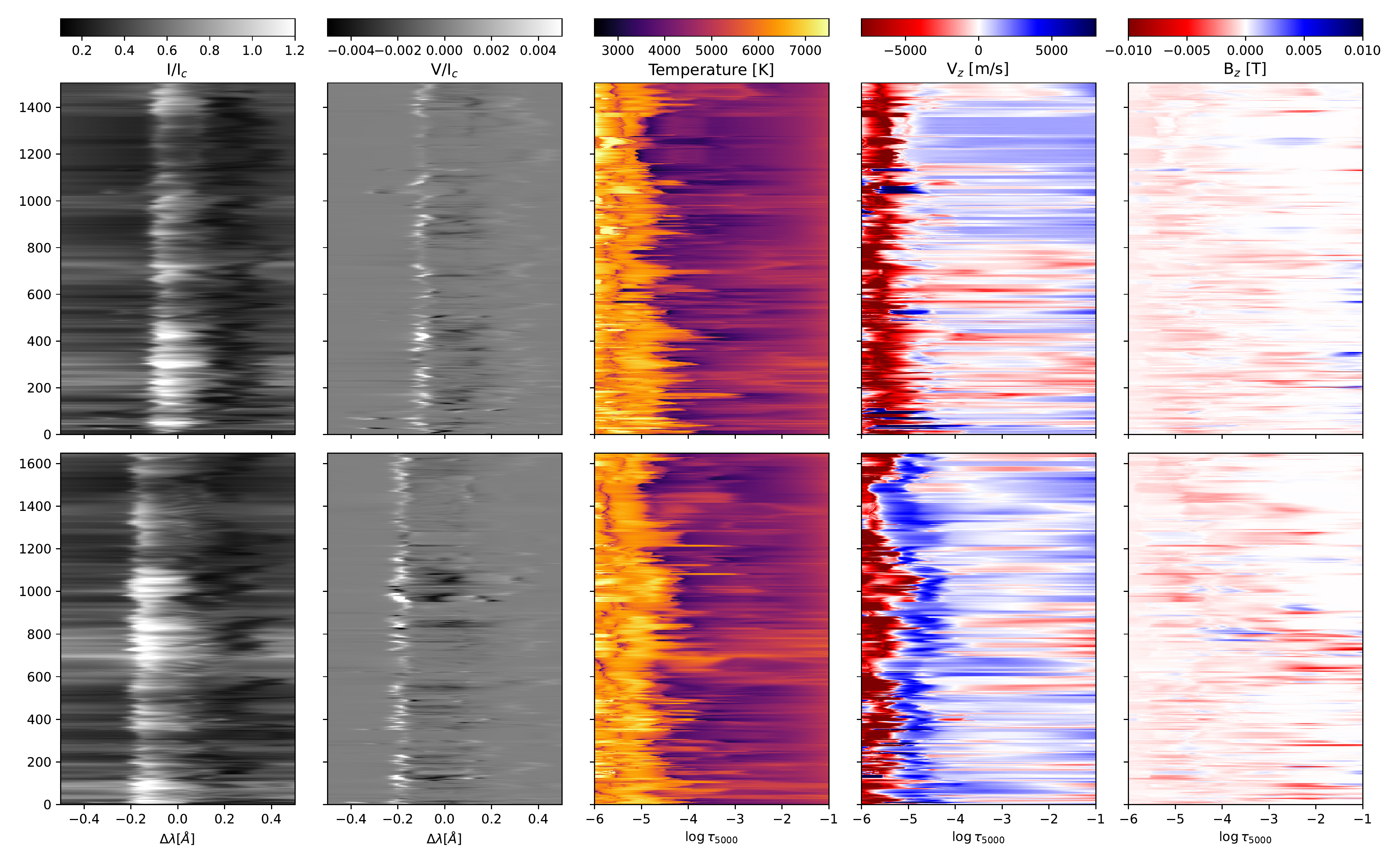}
                        \caption{
                        Stokes $I$ and $V$ profiles for two clusters, along with some atmospheric parameters for their simulation columns. Here showing the $k$-means clusters \#70 (\emph{top row}) and  \#72 (\emph{bottom row}) from Fig. \ref{fig:kmie_clusters}, both of which have CBG-like profiles. All profiles for each cluster are stacked along the vertical axes of the plots, so the y-axis merely counts the profile number. The left column shows the continuum-normalized intensity versus wavelength from line core. The second-from-left column shows the continuum-normalized Stokes $V$ profiles. The last three columns show, respectively, the temperature, the line-of-sight velocity, and the line-of-sight magnetic field strength, as a function of $log(\tau_{5000})$.} 
                        \label{fig:km_lambdalike_logtau}
\end{figure*}

\begin{figure*}
        \centering
                \includegraphics[width=18cm]{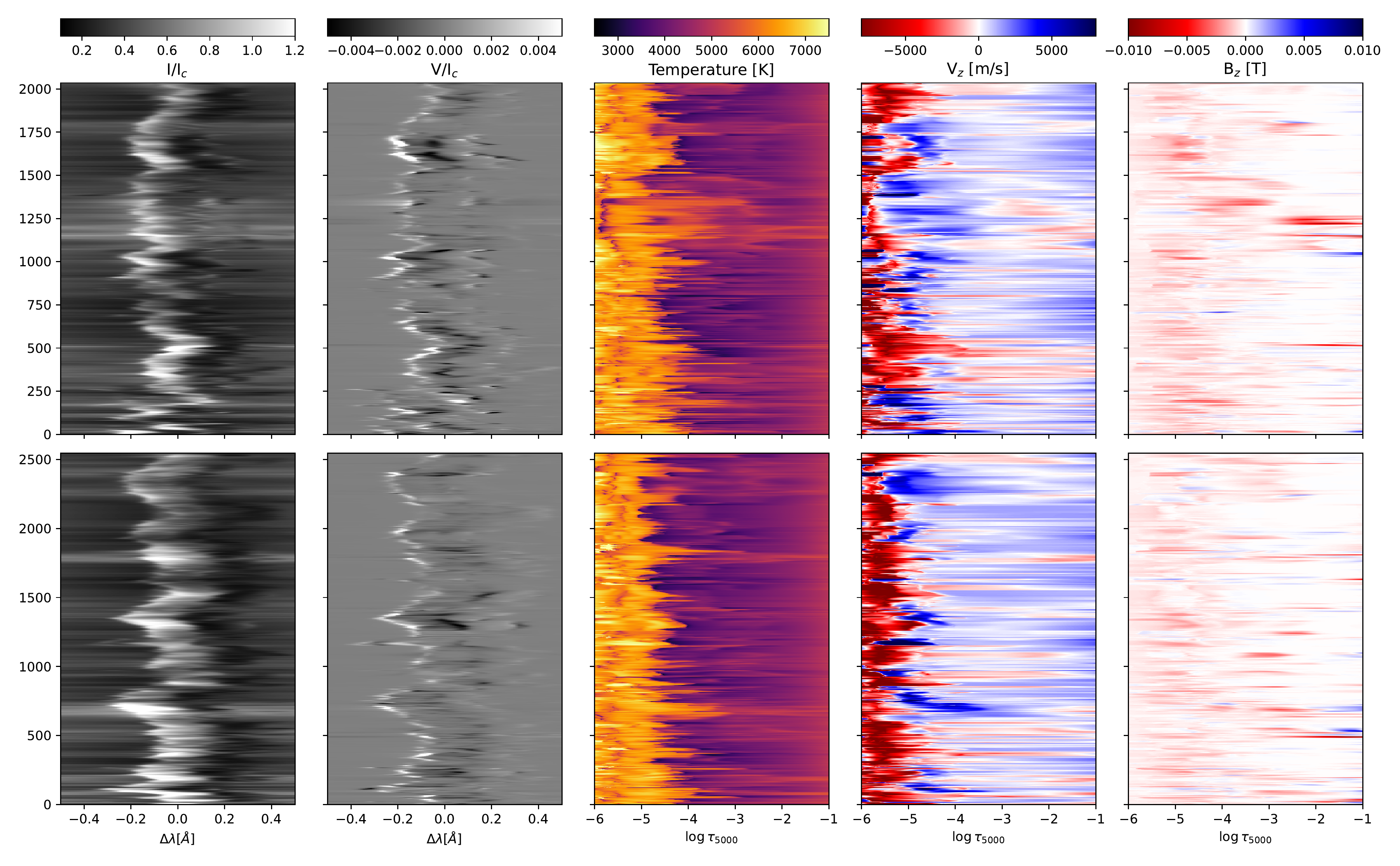}
                        \caption{Same as Fig. \ref{fig:km_lambdalike_logtau}, but for the $k$-Shape clusters \#64 (top) and \#65 (bottom) from Fig. \ref{fig:ksie_clusters}.}
                        \label{fig:ks_lambdalike_logtau}
\end{figure*}

As an example of the sort of analysis facilitated by these kinds of clustering techniques, we decided to perform an in-depth examination of the family of asymmetric blue-lobed single-peaked Stokes I profiles found in Fig. \ref{fig:kmie_clusters} (exemplified by cluster number \#70 and \#72) and Fig. \ref{fig:ksie_clusters} (exemplified by cluster number \#64 and \#65). These profiles are reminiscent of the chromospheric bright grains (CBGs) seen in the \ion{Ca}{II} H and K lines, see for instance \citep{1997ApJ...481..500C, 2022A&A...668A.153M} and references therein, so we call them CBG-like. 

Fig. \ref{fig:km_lambdalike_logtau} shows the Stokes I and Stokes $V$ signals (with each profile normalized to its nearby continuum Stokes I value), as well as the stratification of temperature, line-of-sight velocity, and line-of-sight magnetic field strength for all the profiles belonging to $k$-means cluster \#70 and \#72. The atmospheric quantities are plotted as a function of the logarithm of optical depth for radiation at wavelength 500 nm (5000 Å), $\log(\tau_{5000})$. Throughout this paper we use the convention that positive heights, velocities and vertical magnetic field components point outwards from the solar surface. Each row of Fig. \ref{fig:km_lambdalike_logtau} corresponds to one cluster, and the profiles are stacked along the vertical axis for each panel. The $k$-Shape clusters \#64 and \#65 are shown in a similar fashion in Fig. \ref{fig:ks_lambdalike_logtau}.

Looking at the intensities we see that the clusters are indeed well constrained for the most part. The $k$-means method produces clusters where the emission peak is at approximately the same wavelength throughout each cluster, but with some variance in the other features of the profile shapes. The $k$-Shape method, on the other hand, retrieves clusters where the location of the emission peak varies considerably in wavelength, but the shapes in each cluster seem more consistent in their shapes. For instance, the wavelength distance of the slope from peak to bottom seems to be more regular, and the red-side absorption features show less variance.

As for the Stokes $V$ profiles, with both methods the wavelength positions of the strongest Stokes $V$ signals seem to coincide with the sharpest changes in the intensity as one might expect from the Weak Field Approximation. There does not, however, seem to be any other universal tendencies in Stokes $V$ across all the CBG-like clusters. Similarly for the stratification of the line-of-sight magnetic field strengths, there do not appear to be clear tendencies neither within nor across the clusters. This suggests that the structure of the vertical magnetic field component does not play a direct role in the formation of these CBG-like Stokes I profiles. 

What does seem to be common to all the clusters, and therefore important for the formation of these profile shapes, is the depth-stratification of temperature and line-of-sight velocities. Mostly we see a temperature increase in the atmosphere, followed by a large velocity gradient slightly higher up. Mostly this manifests as upflowing material from below meeting downflowing material from above, but not exclusively as there are some instances of faster downflows from above meeting slower downflows, i.e. there is not necessarily a sign change in the vertical velocity, but there is a significant change in speed. 

That the temperature increase occurs deeper in the atmosphere than the velocity gradient, as well as the fact that the absolute values of the velocity are less important for the formation of these shapes than the presence of a strong gradient, is more easily seen with the $k$-Shape clusters as each of them contains the CBG-like profile shapes at a range of Doppler shifts. In any case, the correlation between the temperature increase, the velocity-gradient and the profile shape is certainly made clearer when comparing the results of both clustering methods.

In terms of explaining the formation of these profiles, we are reminded of the interpretation of \ion{Ca}{II} K and H bright grains provided in \citep{1997ApJ...481..500C} as signatures of acoustic shocks propagating upwards through the chromosphere, with the asymmetry being caused by Doppler shifts of the opacity across the shock front. The increased temperature enhances the local source function, which produces enhanced emission. The velocity gradient to more rapidly downflowing material above the heating event causes an opacity shift as the absorbing material is shifted to redder wavelengths, letting the bluer part of the profile escape while attenuating the redder part. 

A point of note is that the correlation between the atmospheric structure and the CBG-like profile shapes is apparent straight from the clustering when we have access to underlying atmosphere. This allowed a qualitative interpretation of the profiles' formation without having to resort to using response functions or contribution functions, which are ill-defined for the case of 3D radiative transfer.

\subsection{Double peaked profiles}
\label{sec:double_peaked}

\begin{figure*}
        \centering
                \includegraphics[width=18cm]{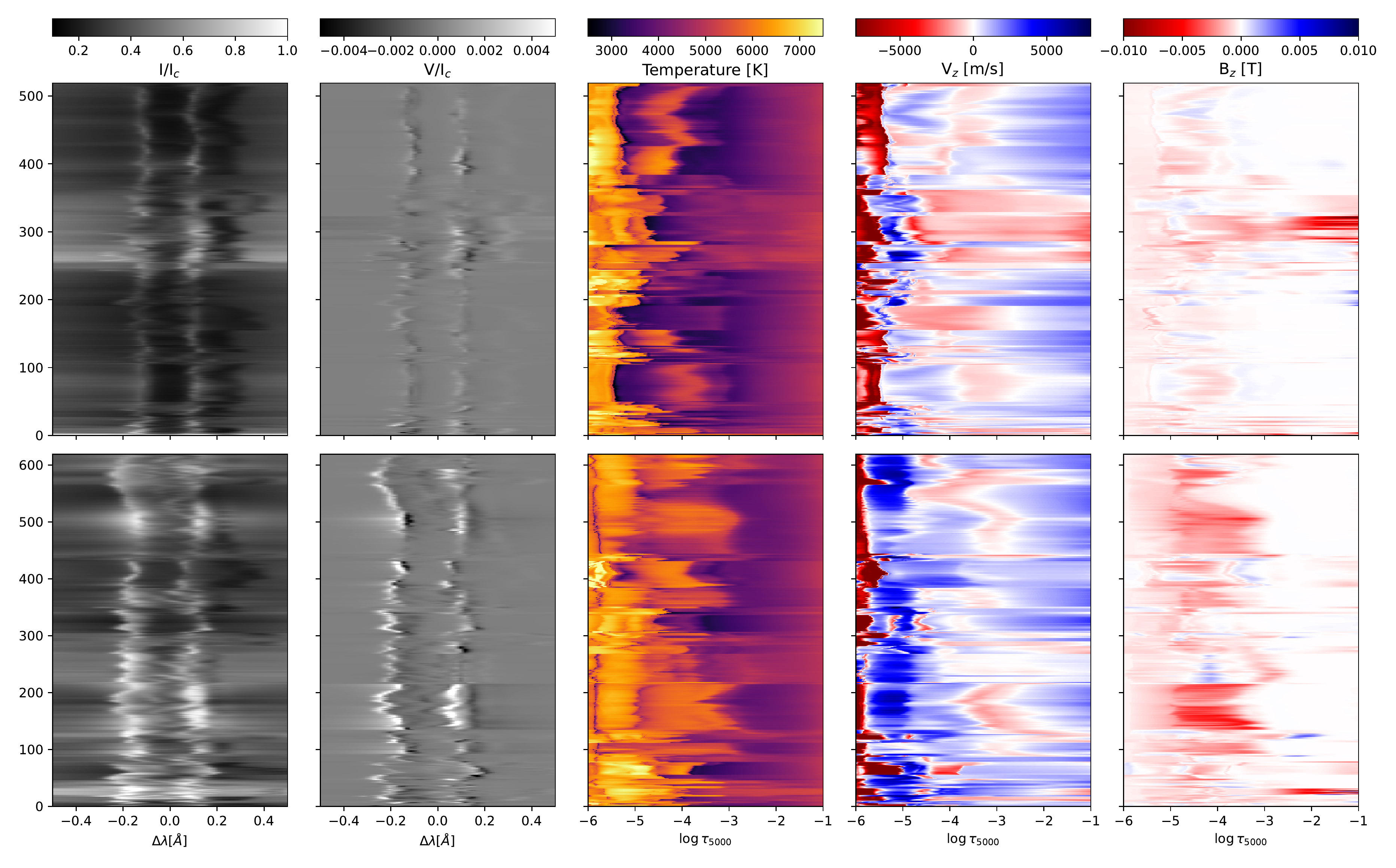}
                        \caption{Same as Fig. \ref{fig:km_lambdalike_logtau}, but for the $k$-means clusters \#98 (top) and \#100 (bottom) from Fig. \ref{fig:kmie_clusters}.} 
                        \label{fig:km_double_peak}
\end{figure*}

\begin{figure*}
        \centering
                \includegraphics[width=18cm]{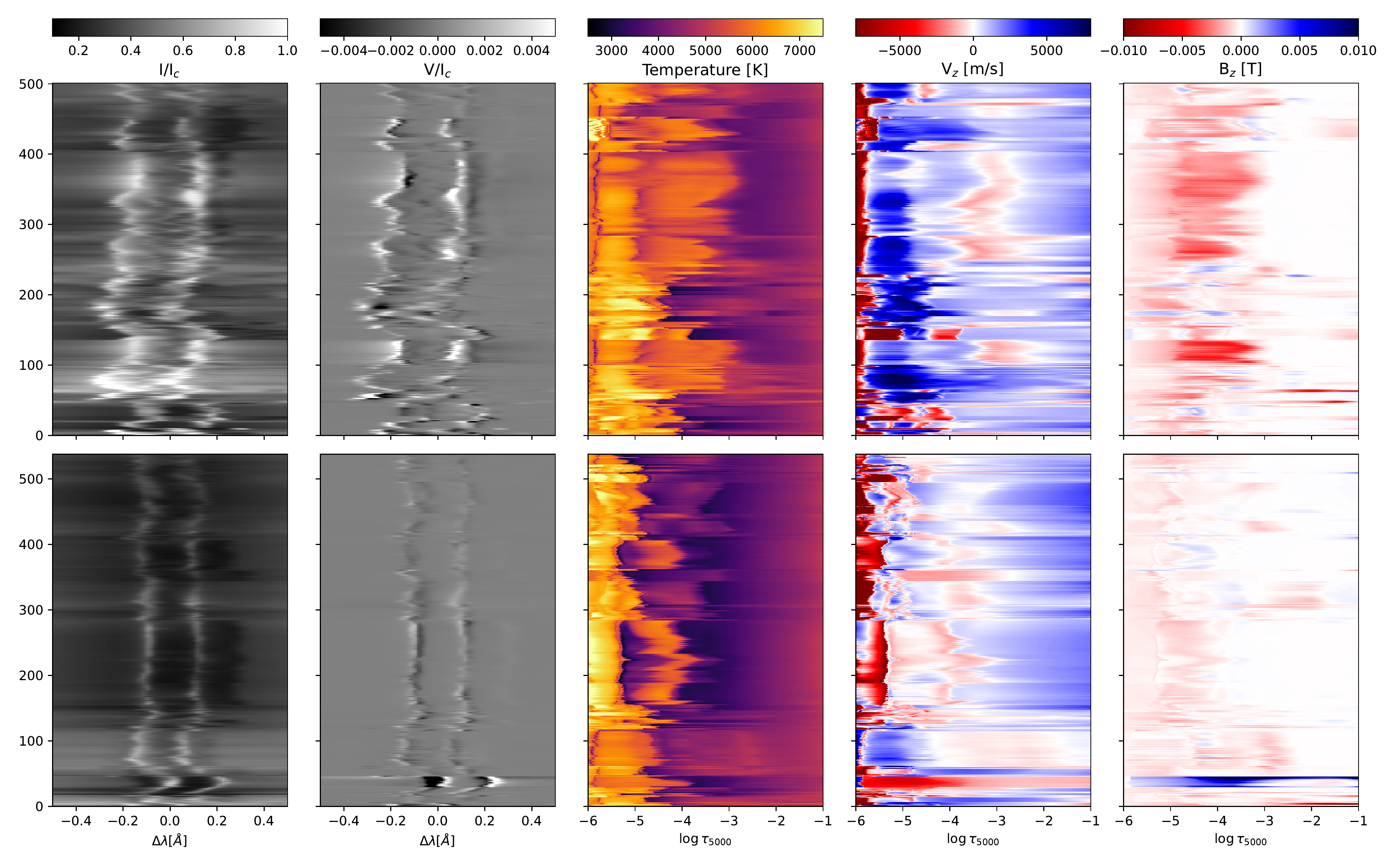}
                        \caption{Same as Fig. \ref{fig:km_lambdalike_logtau}, but for the $k$-Shape clusters \#97 (top) and \#100 (bottom) from Fig. \ref{fig:ksie_clusters}.}
                        \label{fig:ks_double_peak}
\end{figure*}

As another example, we now consider the double peaked profiles seen in $k$-means clusters \#98 and \#100, and in $k$-Shape clusters \#97 and \#100. Similar to Figs. \ref{fig:km_lambdalike_logtau} and \ref{fig:ks_lambdalike_logtau}, the continuum-normalized intensity and Stokes $V$ signals, as well as the height-stratified temperature, line-of-sight velocity, and line-of-sight magnetic field strength for all the individual profiles in each cluster is shown in Figs. \ref{fig:km_double_peak} and \ref{fig:ks_double_peak} for the $k$-means and $k$-Shape clusters, respectively. 

Once again the clusters, on the whole, seem fairly well constrained regarding the shape of the intensity profiles. Here, there seems to be a larger variation in the absolute values of the intensities compared to the previous example. This sort of variation is not unexpected; since the $z$-normalization scales each profile independently to have a standard deviation equal to one our clusters are relatively insensitive to amplitudes, focusing instead on the shapes. Comparing the methods, we see they mostly recover the same profiles. An exception is that the $k$-means cluster \#98 in the top row of Fig. \ref{fig:km_double_peak} has some unique profiles around profile number 300 which appear to have either a very weak left peak or only a single peak on the right, followed by a prominent absorption feature to the right of the rightmost peak. Looking at the temperature and velocity structure for these atypical profiles with suppressed left peaks, it appears they have a temperature enhancement coinciding in height with a moderate downflow. This temperature enhancement persists upwards through a velocity gradient to a region of strong upflow, before it hits a very strong downflow. Their formation can potentially be explained in the same manner as the CBG-like profiles; but with an oppositely signed velocity gradient, and with the strong downflow above the upflow causing the additional strongly redshifted absorption feature.

 Returning to the general behavior of the clusters, we find that the Stokes $V$ profiles seem to behave as expected from the weak-field approximation, in that they follow the behaviour of the intensity profiles. There is, however, a rather interesting region between profile number 200 to 300 in the bottom row of Fig. \ref{fig:km_double_peak}, where the rightmost Stokes $V$ signal is very low despite a gradient in the intensity, and the vertical magnetic field component has a sign change around $\log \tau_{5000} = -4$.

The temperature structure of the atmosphere is more varied for the double peaked profiles, compared to the CBG-like profiles. There are both regions of temperature enhancements with little variation spanning decades in $\log \tau_{5000}$, and hot regions bounded by colder plasma above and below. The common feature for all these double peaked profiles is enhanced temperatures in the range of $-5 < \log \tau_{5000} < -3$. That was also the case for the CBG-like profiles, though the CBG-like profiles seldom showed these colder layers above the first strong temperature increase.

The vertical velocities are also rather varied in their structure, but three general features stand out compared to the CBG-like profiles from before. Firstly, the shift from upflows (or weak downflows) to strong downflows at the top tends to occur at a higher point in the atmosphere. Secondly, the starting points for the temperature enhancements coincide with slower plasma velocities and weaker velocity gradients, as opposed to the CBG-like profiles where the temperature increase starts slightly below strong velocity gradients. Thirdly, we note that the second velocity layer from the top, roughly $-5.5 < \log \tau_{5000} < -4.5$, typically shows low to moderate velocities and fairly modest gradients. As such, the effect of opacity shifting in this layer is less, and both intensity peaks due to the temperature enhancements survive.

Another noteworthy point, is that when these double peaked profiles do have downflows from the top extending deeper (to $\log \tau_{5000}  \approx -5.5$), the downflows are very strong and there is a corresponding absorption feature on the red side of the reddest peak. A possible interpretation is that the previously discussed opacity shifting is so red-shifted in those cases, that it overshoots the red peaks from the slower flowing regions and therefore does not suppress them.

Interestingly, and contrasting with the CBG-like profiles, the vertical component of the magnetic field does in many of these double peaked profiles display some correlations with the vertical velocities and temperature stratifications. To wit, there are areas of Figs. \ref{fig:km_double_peak} and \ref{fig:ks_double_peak} where the velocities change signs coinciding with an appreciable gradient in vertical field strength to more negative (downward) values. Furthermore, the starting heights of the temperature increases coincide with the appearance of the stronger vertical magnetic field components; particularly obvious examples are profiles number 100 through 200 in the bottom row of Fig. \ref{fig:km_double_peak}, and profiles number 300 through 500 in the top row of Fig. \ref{fig:ks_double_peak}. 

In summary, these double peaked profiles seem to arise from a range of different atmospheric conditions. The common features are increased temperatures in the low chromosphere/upper photosphere, coinciding with low or modest velocities and weak velocity gradients.  This, combined with cospatial enhanced vertical magnetic field strengths, suggests that these profiles are not all caused solely by acoustic shocks, in contrast with the CBG-like profiles. Whether the cause of the heating is due to a magnetic phenomenon, or if we simply see already hot plasma being transported, is unclear from this analysis.

\subsection{The strongest Stokes $V$ profiles}
\label{sec:strong_V}

\begin{figure*}
        \centering
                \includegraphics[width=18cm]{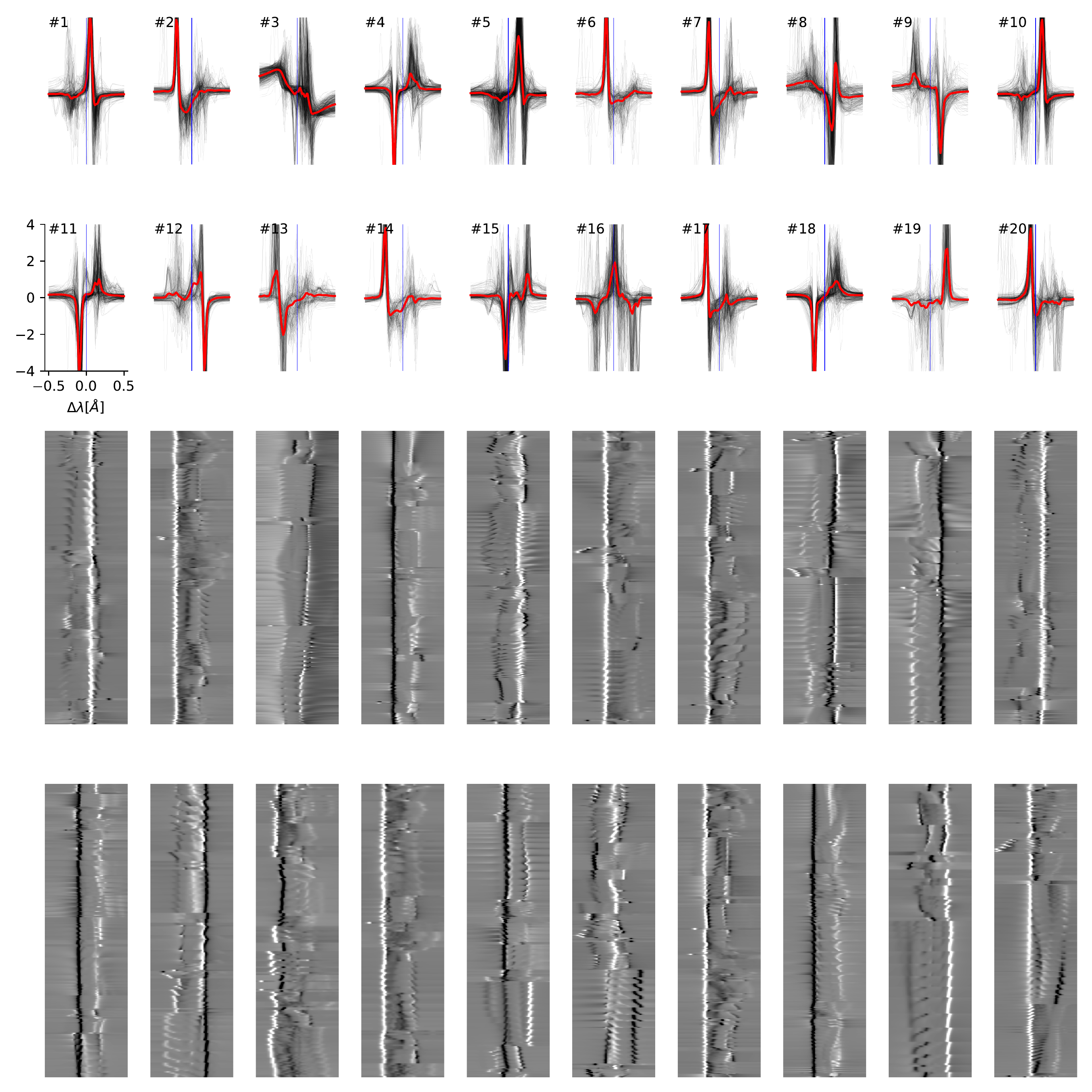}
                        \caption{$k-$means clusters for Stokes $V$ profiles, using 20 clusters on $z$-normalized Stokes $V$. The red line is the cluster centroid (average), and the black lines are all individual profiles assigned to the cluster. The blue line is a visual aid that denotes the position of $\lambda_0$. The bottom two rows show all the $z$-normalized Stokes $V$ profiles belonging to the corresponding clusters in the two top rows, with the individual profiles stacked along the vertical axis. It should be noted that the clusters are not equally populated, so the grey-scale maps will have different densities of profiles along the vertical axis.}
                        \label{fig:kmvestrong_clusters}
\end{figure*}

\begin{figure*}
        \centering
                \includegraphics[width=18cm]{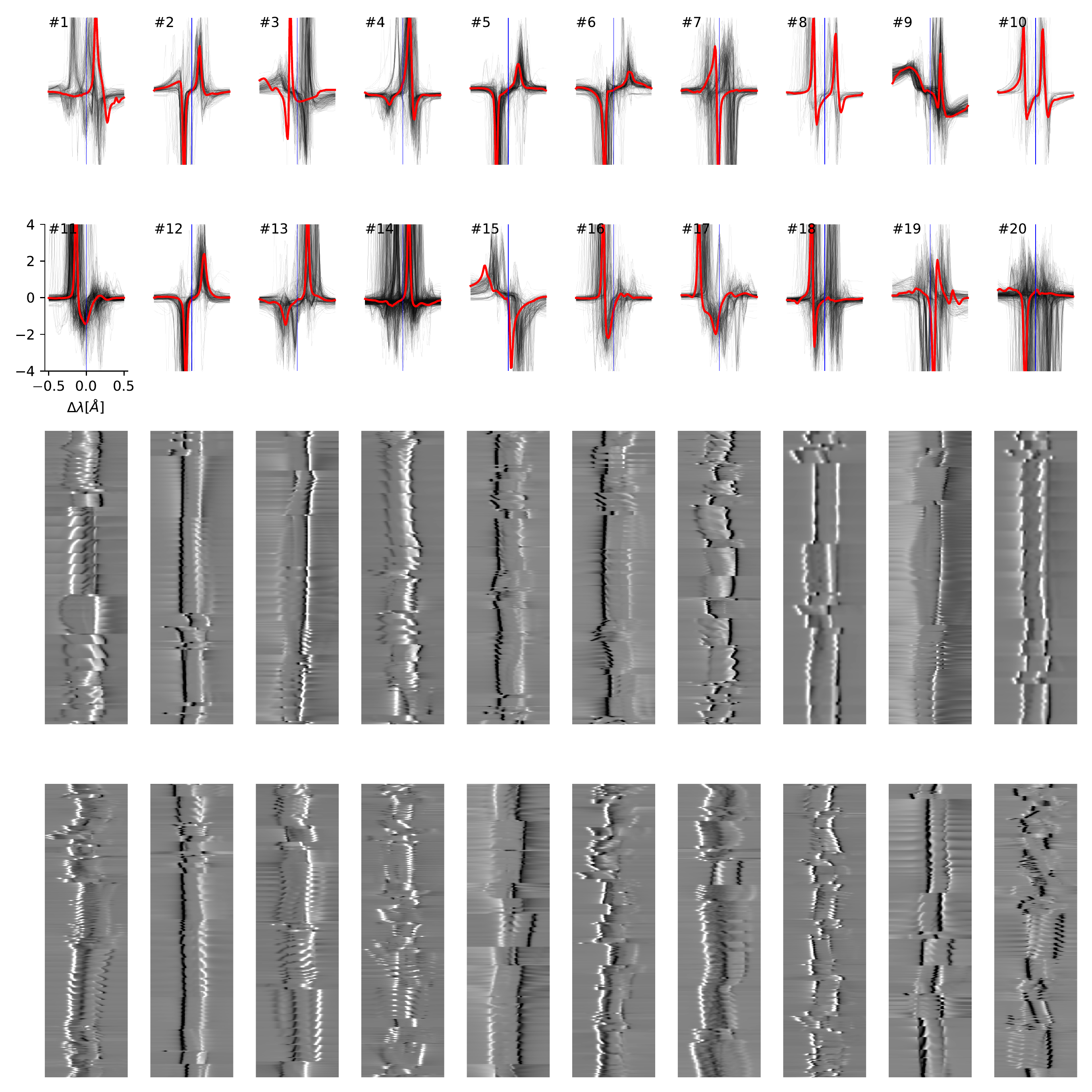}
                        \caption{$k-$shape clusters for Stokes $V$ profiles, using 20 clusters on $z$-normalized Stokes $V$. The red line is the cluster centroid from $k$-Shape, and the black lines are all individual profiles assigned to the cluster. The blue line is a visual aid that denotes the position of $\lambda_0$. The bottom two rows show all the $z$-normalized Stokes $V$ profiles belonging to the corresponding clusters in the two top rows, with the individual profiles stacked along the vertical axis. It should be noted that the clusters are not equally populated, so the grey-scale maps will have different densities of profiles along the vertical axis.}
                        \label{fig:ksvestrong_clusters}
\end{figure*}

We have so far focused on the clustering of intensity profiles, since the $z$-normalization scaled Stokes $V$ signals of very different amplitudes to a misleadingly similar range. Many of our Stokes $V$ profiles contained only very weak signals, and clustering according to the shapes of such weak signals should not be expected to provide much diagnostic information. However, by restricting ourselves to look only at the Stokes $V$ profiles containing an (unsigned) amplitude larger than $0.5\%$ of the nearby continuum intensity we could perform a clustering on profiles with similar strengths. Out of our $512 \times 512$ synthetic profiles, only 7054 ($\approx 2.7\%$) matched that selection criterion. The results of $k$-means and $k$-Shape clustering with $k=20$ clusters on this subset of Stokes $V$ profiles are shown in Fig. \ref{fig:kmvestrong_clusters} and Fig. \ref{fig:ksvestrong_clusters} respectively. 

In this case, we deliberately selected a rather low number of clusters. This was partly done to avoid having clusters with very few members considering our reduced dataset, and partly to compare the performance of the two methods when using a very limited, and possibly too low, number of clusters. It is obvious from looking at Figs. \ref{fig:kmvestrong_clusters} and \ref{fig:ksvestrong_clusters} that 20 clusters is not sufficient to capture all the complexities present in the profiles with either method, though the clusterings do reproduce the primary features of the profiles. 

Comparing these two clustering results reveals some interesting differences. Most noticeably, not all shapes are common to both methods. The double peaked Stokes V profiles of cluster number \#8 and \#10 in the $k$-Shape result are not retrieved as a separate class by the $k$-means method; instead they are mixed into most of the $k$-means clusters, though primarily into \#1, \#7, \#10, \#12 and \#17. On the other hand, the valley-peak-valley shape apparent in cluster number \#16 from the $k$-means method does not appear in the $k$-Shape case. Looking in more detail at the individual profiles comprising that cluster, we find almost no profiles with a shape similar to that of the cluster mean. The triple-lobed shape of the cluster mean (marked in red) is instead mostly a mix of valley-peak and peak-valley shapes. In this case, the $k$-Shape centroids are more faithful representations of the shapes picked up by each cluster.

In general, the clusters found by $k$-means contain one dominant feature, like a peak, a dip, or both, at a certain wavelength position with considerable variation in the rest of the signal. Furthermore, looking at cluster \#13 or \#16 in Fig. \ref{fig:kmvestrong_clusters} we see that when the dominant feature in the cluster is multi-lobed, it might actually be a mix of single-lobed and multi-lobed signals grouped together, so long as their lobes occur at the same wavelength. This type of shape-mixing does not happen as readily with $k$-Shapes, contrast $k$-means cluster \#13 with $k$-Shape cluster \#15 and \#17. Also, $k$-Shape seems to retrieve profiles with more commonality also at the weaker parts of the signal; compare for instance $k$-means clusters  \#5,  \#10 and \#19 with $k$-Shape clusters \#1, \#5 and \#13. $k$-Shape does, however, occasionally struggle when excessive shifts of the signal causes clipping of the features at the edges, which can be most easily seen in cluster \#1, \#19 or \#20 of Fig. \ref{fig:ksvestrong_clusters}. While it is by no means perfect, we find, in conclusion, that $k$-Shapes performs markedly better than $k$-means at identifying shapes with this particular combination of complex signals and low number of clusters. How well that observation generalizes to other datasets, or cluster numbers, or both, is not clear, and beyond the scope of the current work. It does, however, indicate the type of problems where $k$-Shape can potentially provide an advantage over $k$-means. As a note, we have also performed this clustering experiment with $k$-means on the continuum-normalized Stokes V profiles and found that their behavior is very similar to the $z$-normalized case discussed above. 

\section{Discussion and Conclusions}
\label{sec:conclusion}

We have used the $k$-means and $k$-Shape clustering techniques to group according to profile shape synthetic \ion{Ca}{II} intensity and Stokes $V$ profiles, generated by 3D radiative transfer calculations from a 3D MHD simulation. 

Using $k=100$ clusters for the intensities resulted in both methods retrieving qualitatively similar `families' of clusters. While the $k$-means method produced clusters whose features were strongly coherent with regard to wavelength, the $k$-Shape method, being insensitive to Doppler shifts, produced clusters where the same shape appeared over a range of wavelength shifts. Regarding the methods' shortcomings, we found that $k$-Shape occasionally would mislabel some profiles by clipping the signals at the edges when comparing across Doppler shifts, while $k$-means at times would lump rather differently shaped profiles together so long as their strongest feature occurred at the same wavelength.

Armed with full knowledge of the simulation's atmospheric parameters, we took an in-depth look at a particular set of profile shapes and arrived at an explanation of their formation by looking at the correlations in the underlying atmospheric structure. We remark that the most interesting aspect of this exercise was not the description itself of how those profile shapes are formed, but rather how we arrived at it. In that use case, there did not appear to be much benefit in using one method over the other in terms of the results; though $k$-means was significantly quicker computationally. However, we do note that using both methods gave a stereoscopic view of the data, making it easier to determine which atmospheric quantities were important. 

Doing a clustering analysis of the Stokes $V$ profiles, based on their shapes, proved difficult due to the large variations in signal strength being masked by the $z$-normalization required by $k$-Shape, causing strong and weak signals to appear deceivingly similar. Restricting ourselves to a subset of the strongest Stokes $V$ profiles, we performed a clustering with $k=20$ clusters using both methods. We found that the methods showed the same tendencies as with the intensity, but more strongly pronounced due to the lower number of clusters and more complex shapes. In this setting we found that $k$-means clearly performed qualitatively worse than $k$-Shape at creating clusters with coherent shapes; though is difficult to quantitatively compare the methods since they use very different metrics.

In conclusion, $k$-Shape seems interesting for use cases where one wants human interpretation and small numbers of clusters. Another interesting possibility is to use the $k$-Shape distance metric to search an observation or simulation for the profiles with shape most similar to a certain prototype, for example when trying to detect Ellerman bombs. We want to stress that $k$-Shape is, however, not at all suited to usage cases like \citep{2019ApJ...875L..18S}, where the purpose of clustering is to speed up inversions, as the centroids found by $k$-Shape do not correspond  to a definite Doppler-shift nor to an absolute intensity. In those cases, $k$-means is the better option, and one can easily increase the number of clusters beyond what a human can reasonably process. For a qualitative clustering, aimed towards human interpretation and with a comparatively small number of clusters, we find that $k$-Shape can be a useful complement to, and sometimes better than, the more well-known $k$-means method.

\begin{acknowledgements}
The authors wish to thank Mats Carlsson for providing the Bifrost atmosphere used in this paper. We also wish to thank the anonymous referee for comments and suggestions that improved the clarity of this manuscript. This work has been supported by the Research Council of Norway through its Centers of Excellence scheme, project number 262622. Computational resources have been provided by UNINETT Sigma2 - the National Infrastructure for High Performance Computing and Data Storage in Norway. The computations were enabled by resources provided by the Swedish National Infrastructure for Computing (SNIC) at the PDC Centre for High Performance Computing (PDC-HPC) at the Royal Institute of Technology partially funded by the Swedish Research Council through grant agreement no. 2018-05973.
\end{acknowledgements}
\bibliography{Refs.bib}

\begin{thebibliography}{30}
\expandafter\ifx\csname natexlab\endcsname\relax\def\natexlab#1{#1}\fi

\bibitem[{Arthur \& Vassilvitskii(2007)}]{10.5555/1283383.1283494}
Arthur, D. \& Vassilvitskii, S. 2007, in Proceedings of the Eighteenth Annual
  ACM-SIAM Symposium on Discrete Algorithms, SODA '07 (USA: Society for
  Industrial and Applied Mathematics), 1027–1035

\bibitem[{{Barczynski} {et~al.}(2021){Barczynski}, {Harra}, {Kleint}, {Panos},
  \& {Brooks}}]{2021A&A...651A.112B}
{Barczynski}, K., {Harra}, L., {Kleint}, L., {Panos}, B., \& {Brooks}, D.~H.
  2021, \aap, 651, A112

\bibitem[{{Bose} {et~al.}(2019){Bose}, {Henriques}, {Joshi}, \& {Rouppe van der
  Voort}}]{2019A&A...631L...5B}
{Bose}, S., {Henriques}, V. M.~J., {Joshi}, J., \& {Rouppe van der Voort}, L.
  2019, \aap, 631, L5

\bibitem[{{Bose} {et~al.}(2021){Bose}, {Joshi}, {Henriques}, \& {Rouppe van der
  Voort}}]{2021A&A...647A.147B}
{Bose}, S., {Joshi}, J., {Henriques}, V. M.~J., \& {Rouppe van der Voort}, L.
  2021, \aap, 647, A147

\bibitem[{{Carlsson} \& {Stein}(1997)}]{1997ApJ...481..500C}
{Carlsson}, M. \& {Stein}, R.~F. 1997, \apj, 481, 500

\bibitem[{{Gudiksen} {et~al.}(2011){Gudiksen}, {Carlsson}, {Hansteen}, {Hayek},
  {Leenaarts}, \& {Mart{\'\i}nez-Sykora}}]{2011A&A...531A.154G}
{Gudiksen}, B.~V., {Carlsson}, M., {Hansteen}, V.~H., {et~al.} 2011, \aap, 531,
  A154

\bibitem[{{Joshi} {et~al.}(2020){Joshi}, {Rouppe van der Voort}, \& {de la Cruz
  Rodr{\'\i}guez}}]{2020A&A...641L...5J}
{Joshi}, J., {Rouppe van der Voort}, L. H.~M., \& {de la Cruz Rodr{\'\i}guez},
  J. 2020, \aap, 641, L5

\bibitem[{{Khomenko} {et~al.}(2005){Khomenko}, {Shelyag}, {Solanki}, \&
  {V{\"o}gler}}]{2005A&A...442.1059K}
{Khomenko}, E.~V., {Shelyag}, S., {Solanki}, S.~K., \& {V{\"o}gler}, A. 2005,
  \aap, 442, 1059

\bibitem[{{Kleint} \& {Panos}(2022)}]{2022A&A...657A.132K}
{Kleint}, L. \& {Panos}, B. 2022, \aap, 657, A132

\bibitem[{{Kuckein} {et~al.}(2020){Kuckein}, {Gonz{\'a}lez Manrique}, {Kleint},
  \& {Asensio Ramos}}]{2020A&A...640A..71K}
{Kuckein}, C., {Gonz{\'a}lez Manrique}, S.~J., {Kleint}, L., \& {Asensio
  Ramos}, A. 2020, \aap, 640, A71

\bibitem[{{Leenaarts} \& {Carlsson}(2009)}]{2009ASPC..415...87L}
{Leenaarts}, J. \& {Carlsson}, M. 2009, in Astronomical Society of the Pacific
  Conference Series, Vol. 415, The Second Hinode Science Meeting: Beyond
  Discovery-Toward Understanding, ed. B.~{Lites}, M.~{Cheung}, T.~{Magara},
  J.~{Mariska}, \& K.~{Reeves}, 87

\bibitem[{{Leenaarts} {et~al.}(2013{\natexlab{a}}){Leenaarts}, {Pereira},
  {Carlsson}, {Uitenbroek}, \& {De Pontieu}}]{2013ApJ...772...89L}
{Leenaarts}, J., {Pereira}, T.~M.~D., {Carlsson}, M., {Uitenbroek}, H., \& {De
  Pontieu}, B. 2013{\natexlab{a}}, \apj, 772, 89

\bibitem[{{Leenaarts} {et~al.}(2013{\natexlab{b}}){Leenaarts}, {Pereira},
  {Carlsson}, {Uitenbroek}, \& {De Pontieu}}]{2013ApJ...772...90L}
{Leenaarts}, J., {Pereira}, T.~M.~D., {Carlsson}, M., {Uitenbroek}, H., \& {De
  Pontieu}, B. 2013{\natexlab{b}}, \apj, 772, 90

\bibitem[{MacQueen(1967)}]{1967_Macqueen}
MacQueen, J. 1967, in Proc. of the fifth Berkeley Symposium on Mathematical
  Statistics and Probability, ed. L.~M. Le~Cam \& J.~Neyman, Vol.~1 (University
  of California Press), 281--297

\bibitem[{{Mathur} {et~al.}(2022){Mathur}, {Joshi}, {Nagaraju}, {Rouppe van der
  Voort}, \& {Bose}}]{2022A&A...668A.153M}
{Mathur}, H., {Joshi}, J., {Nagaraju}, K., {Rouppe van der Voort}, L., \&
  {Bose}, S. 2022, \aap, 668, A153

\bibitem[{{Moe} {et~al.}(2022){Moe}, {Pereira}, \&
  {Carlsson}}]{2022A&A...662A..80M}
{Moe}, T.~E., {Pereira}, T. M.~D., \& {Carlsson}, M. 2022, \aap, 662, A80

\bibitem[{{N{\'o}brega-Siverio} {et~al.}(2021){N{\'o}brega-Siverio},
  {Guglielmino}, \& {Sainz Dalda}}]{2021A&A...655A..28N}
{N{\'o}brega-Siverio}, D., {Guglielmino}, S.~L., \& {Sainz Dalda}, A. 2021,
  \aap, 655, A28

\bibitem[{{Panos} {et~al.}(2018){Panos}, {Kleint}, {Huwyler}, {Krucker},
  {Melchior}, {Ullmann}, \& {Voloshynovskiy}}]{2018ApJ...861...62P}
{Panos}, B., {Kleint}, L., {Huwyler}, C., {et~al.} 2018, \apj, 861, 62

\bibitem[{Paparrizos \& Gravano(2015)}]{10.1145/2723372.2737793}
Paparrizos, J. \& Gravano, L. 2015, in Proceedings of the 2015 ACM SIGMOD
  International Conference on Management of Data, SIGMOD '15 (New York, NY,
  USA: Association for Computing Machinery), 1855–1870

\bibitem[{Pedregosa {et~al.}(2011)Pedregosa, Varoquaux, Gramfort, Michel,
  Thirion, Grisel, Blondel, Prettenhofer, Weiss, Dubourg, Vanderplas, Passos,
  Cournapeau, Brucher, Perrot, \& Duchesnay}]{scikit-learn}
Pedregosa, F., Varoquaux, G., Gramfort, A., {et~al.} 2011, Journal of Machine
  Learning Research, 12, 2825

\bibitem[{{Pereira} {et~al.}(2013){Pereira}, {Leenaarts}, {De Pontieu},
  {Carlsson}, \& {Uitenbroek}}]{2013ApJ...778..143P}
{Pereira}, T.~M.~D., {Leenaarts}, J., {De Pontieu}, B., {Carlsson}, M., \&
  {Uitenbroek}, H. 2013, \apj, 778, 143

\bibitem[{{Pietarila} {et~al.}(2007){Pietarila}, {Socas-Navarro}, \&
  {Bogdan}}]{2007ApJ...663.1386P}
{Pietarila}, A., {Socas-Navarro}, H., \& {Bogdan}, T. 2007, \apj, 663, 1386

\bibitem[{{Sainz Dalda} {et~al.}(2022){Sainz Dalda}, {Agrawal}, {De Pontieu},
  \& {Gosic}}]{2022arXiv221109103S}
{Sainz Dalda}, A., {Agrawal}, A., {De Pontieu}, B., \& {Gosic}, M. 2022, arXiv
  e-prints, arXiv:2211.09103

\bibitem[{{Sainz Dalda} {et~al.}(2019){Sainz Dalda}, {de la Cruz
  Rodr{\'\i}guez}, {De Pontieu}, \& {Go{\v{s}}i{\'c}}}]{2019ApJ...875L..18S}
{Sainz Dalda}, A., {de la Cruz Rodr{\'\i}guez}, J., {De Pontieu}, B., \&
  {Go{\v{s}}i{\'c}}, M. 2019, \apjl, 875, L18

\bibitem[{{S{\'a}nchez Almeida} \& {Lites}(2000)}]{2000ApJ...532.1215S}
{S{\'a}nchez Almeida}, J. \& {Lites}, B.~W. 2000, \apj, 532, 1215

\bibitem[{Steinhaus(1956)}]{1956_Steinhaus}
Steinhaus, H. 1956, Bulletin de l’Acad\'emie Polonaise des Sciences, Cl.
  {III} --- Vol. {IV}, 801

\bibitem[{Tavenard {et~al.}(2020)Tavenard, Faouzi, Vandewiele, Divo, Androz,
  Holtz, Payne, Yurchak, Ru{\ss}wurm, Kolar, \& Woods}]{JMLR:v21:20-091}
Tavenard, R., Faouzi, J., Vandewiele, G., {et~al.} 2020, Journal of Machine
  Learning Research, 21, 1

\bibitem[{{Verma} {et~al.}(2021){Verma}, {Matijevi{\v{c}}}, {Denker},
  {Diercke}, {Dineva}, {Balthasar}, {Kamlah}, {Kontogiannis}, {Kuckein}, \&
  {Pal}}]{2021ApJ...907...54V}
{Verma}, M., {Matijevi{\v{c}}}, G., {Denker}, C., {et~al.} 2021, \apj, 907, 54

\bibitem[{{Viticchi{\'e}} \& {S{\'a}nchez Almeida}(2011)}]{2011A&A...530A..14V}
{Viticchi{\'e}}, B. \& {S{\'a}nchez Almeida}, J. 2011, \aap, 530, A14

\bibitem[{{Woods} {et~al.}(2021){Woods}, {Sainz Dalda}, \& {De
  Pontieu}}]{2021ApJ...922..137W}
{Woods}, M.~M., {Sainz Dalda}, A., \& {De Pontieu}, B. 2021, \apj, 922, 137

\end{thebibliography}

\end{document}